\documentclass[aps,prl,twocolumn,showpacs]{revtex4-1}
\usepackage{amsmath,amsfonts,bm,color,amssymb}
\usepackage{graphicx,epsfig,overpic,graphics}

\begin{document}

\title{Mutual Diffusion of Inclusions in Freely-Suspended Smectic Liquid Crystal Films}

\date{\today}
\author{Zhiyuan Qi}
\author{Zoom Hoang Nguyen}
\author{Cheol Soo Park}
\author{Matthew A.~Glaser}
\author{Joseph E.~Maclennan}
\author{Noel A.~Clark}
\affiliation{Department of Physics and the Liquid Crystal Materials Research Center, University of Colorado, Boulder, Colorado, 80309, USA}

\author{Tatiana Kuriabova}
\author{Thomas R.~Powers}
\affiliation{School of Engineering, Brown University, Providence, Rhode Island, 02912, USA}
\affiliation{Department of Physics, Brown University, Providence, Rhode Island, 02912, USA}

\date{\today}

\begin{abstract}
We study experimentally and theoretically the hydrodynamic interaction of pairs of circular inclusions in two-dimensional, fluid smectic membranes suspended  in air. By analyzing their Brownian motion, we find that the radial mutual mobilities of identical inclusions are independent of their size but that the angular coupling becomes strongly size-dependent when their radius exceeds a characteristic hydrodynamic length. The observed dependence of the mutual mobilities on inclusion size is described well for arbitrary separations by a model that generalizes the Levine/MacKintosh theory of point-force response functions and uses a boundary-element approach to calculate the mobility matrix.
\end{abstract}

\maketitle

Since many of the processes critical for the life of the cell take place in the plasma membrane or in the membranes of organelles, the physics of transport, diffusion, and aggregation of particles in thin, fluid membranes is of fundamental interest.
A distinctive feature of the hydrodynamic interactions between particles embedded in such a membrane is that they are, in general, mediated both by the membrane and by the surrounding fluid. Saffman and Delbr\"uck  (SD) studied theoretically the mobility of a single particle of radius $a$ in a fluid layer of viscosity $\eta$ and thickness $h$ embedded in a different fluid of viscosity $\eta'$, and found that the mobility of the inclusions depends on the Saffman length $\ell_S=\eta h/(2\eta')$~\cite{SaffmanDelbruck1975}. The general dependence of mobility on size and its crossover from three-dimensional to two-dimensional behavior as the particle size is reduced was subsequently calculated theoretically by Hughes, Pailthorpe and White (HPW)~\cite{HPW1981}. For large particles with $a\gg\ell_S$, the mobility scales as $1/(\eta' a)$, just as for a particle in a three-dimensional fluid, while for small particles with $a\ll\ell_S$, the mobility scales as $\log(2\ell_S/a)/(\eta h)$, the SD result. This crossover behavior was recently demonstrated experimentally for isolated inclusions in thin, fluid smectic films by Nguyen~{\em et al.}~\cite{Nguyen2010}.

The hydrodynamic interactions between several inclusions in a fluid membrane are complicated and are much less well studied. When the membrane has multiple inclusions, the mobilities of the inclusions depend not only on their size and the drag from the surrounding fluid(s) but also on the hydrodynamic interactions between them. Bussell~{\em et al.}~\cite{Hammer1992} extended the SD theory in order to calculate the mobilities of two cylinders in a membrane in the limit $a\ll\ell_S$. Levine and MacKintosh, in a study of the microrheology of viscoelastic membranes~\cite{Levine_MacKintosh2002}, derived the response function for a point force in a two-dimensional (2D) fluid, an approach to computing the mobility matrix in the far-field limit that forms the basis of our generalized model of the hydrodynamic interactions of multiple inclusions presented below.
Several relevant experiments have been carried out. For example, Cheung~{\em et al.} measured the diffusion of colloidal particles embedded in soap films and showed that their long-range hydrodynamic interactions were $2$D in nature~\cite{Cheung1996}. Di~Leonardo~{\em et al.} used laser tweezers to manipulate colloidal particles in thick soap films in order to determine their mutual $2$D eigenmobilities in the limit of long Saffman lengths~\cite{Leonardo2008}. In an earlier study, Prasad~{\em et al.} probed experimentally the correlated motion of colloidal particles at the air-water interface~\cite{Prasad2006}. In their analysis of the crossover of the correlated motion from 2D interface-dominated behavior at high surface viscosities to bulk fluid-dependent behavior at lower surface viscosities, they assumed that the particles were sufficiently dilute that they could be treated as points. We will show that this far-field approximation is not universally valid, however, for inclusions that are close together, a common situation, for example, in cellular membranes.

\begin{figure}[tbh]
\includegraphics[width=8cm]{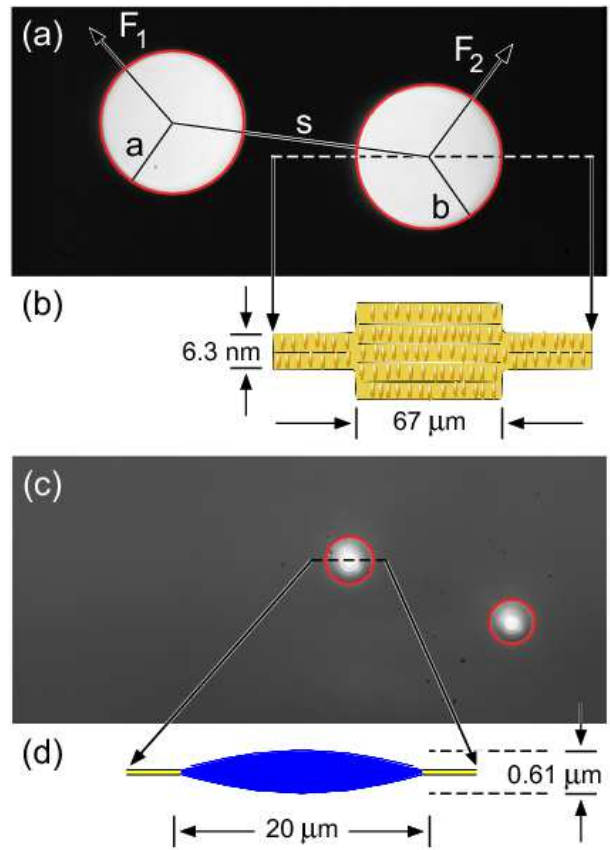}  
\caption{(Color online) Island and silicone oil droplet pairs in thin 8CB films viewed in reflection. (a) Pair of islands with radii $a$ and $b$ and separation $s$ subject to forces $F_1$ and $F_2$. (b) Schematic cross-section of a five-layer island in a two-layer film (not to scale).  (c) Pair of silicone oil droplets in a six-layer film. (d) Cross-section of a typical silicone oil inclusion measured using optical interference in monochromatic light (not to scale).}
\label{fig:island}
\end{figure}

In this Letter, we report measurements and theoretical modeling of the mobilities of pairs of inclusions in smectic~A liquid crystal films that are only a few nanometers thick. These films are extremely thin and stable ~\cite{Young1978}, and provide an ideal platform for studying hydrodynamics in 2D fluids~\cite{Muzny1992,Geminard1997,Schneider2006,Eremin2011}.
In our experiments, we observe the spontaneous Brownian motion of silicone oil droplets and smectic ``islands'' embedded in the liquid crystal film, shown in Fig.~\ref{fig:island}. The islands are disk-shaped, thicker regions of the film bounded by edge dislocations (Fig.~\ref{fig:island}a,b) that can be created with diameters between a few and several hundred $\mu {\rm m}$~\cite{Nguyen2010}. The oil droplets form lens-shaped rather than flat inclusions (Fig.~\ref{fig:island}c,d). A motivation for using oil droplets in addition to islands is that they are essentially insoluble in liquid crystal and their sizes remain essentially constant over long time intervals. This is particularly useful for making mobility measurements of very small inclusions because smectic islands $\alt 10 \;\mu\mathrm{m}$ in diameter shrink rapidly and vanish within a few minutes in our experiment, whereas silicone oil droplets of similar size remain practically unchanged for half an hour or more.

The liquid crystal material used in our experiment is 8CB (4\ensuremath{'}-n-octyl-4\ensuremath{'}-cyanobiphenyl, Sigma-Aldrich), which is in the smectic A phase at room temperature. The density and viscosity of 8CB are $\rho \approx 0.96 \; \mathrm{g/cm}^3$~\cite{Leadbetter1976} and $\eta = 0.052~{\rm Pa} \cdot {\rm s}$~\cite{Schneider2006} respectively, while the viscosity of ambient air is $\eta'=1.827\times 10^{-5}~{\rm Pa} \cdot {\rm s}$~\cite{Robert1973}. Each molecular layer has a thickness of $3.17\,{\rm nm}$~\cite{Davidov1979}. Freely-suspended films from two to six layers thick , corresponding to Saffman lengths between $9$ and $27 \;\mu\mathrm{m}$, were formed by spreading a small amount of the liquid crystal across a $5\,{\rm mm}$-diameter hole in a glass cover slip and were then observed using reflected light microscopy. Immediately after a film is drawn, one typically observes many islands, with a range of diameters. The film can then be gently sheared using an air jet to break larger islands into smaller ones.
To create oil droplets, the LC film was put into a vacuum chamber, where it was left until it was uniform in thickness.  A double-sealed rotary pump (Welch Model~$1400$) was then used to reduce the pressure to $3 \times 10^{-3}$~Torr. After about an hour, a small amount of vaporized pump oil has made its way to the film chamber, where it eventually condenses onto the film and forms visible droplets with diameters from about $4$ to $15\, \mu {\rm m}$. We then restore the chamber to atmospheric pressure and observe the diffusion of the droplets. Isolated pairs of islands (or oil droplets) of similar sizes and far from other inclusions and the film boundaries were selected for study.

The film is carefully leveled to minimize any gravitational drift, allowing us to record, with high spatial resolution and a typical video frame rate of 24 fps, the motion of inclusion pairs for several minutes before they diffuse out of the field of view. We use Canny's method for edge detection~\cite{canny1986} and Taubin's curve fitting algorithm ~\cite{Taubin1991} to measure accurately the positions and sizes of the inclusions in each frame. The effects of any overall drift are removed analytically from the resulting trajectories~\cite{Nguyen2010}. The film thickness, an integral number of smectic layers, is determined precisely by comparing the reflectivities of the film and black glass~\cite{Rosenblatt1980}.

In order to quantify the effects of the long-range hydrodynamic interactions between the inclusions,
we consider two circular domains of radii $a$ and $b$ that are subjected to external forces $\mathbf{F}_1$ and $\mathbf{F}_2$, respectively. Since we are in the low Reynolds number regime, inertial effects are unimportant and the hydrodynamics can be described using the Stokes equations. (For even the most rapidly moving inclusions in our experiments, where $v \sim 1 \; \mathrm{\mu m/s}$, both $Re=\rho v a/\eta \ll 1$ and $Re^\prime=\rho^\prime v a/\eta^\prime \ll 1$, where $\rho^\prime$ is the density of air.) The linearity of these equations implies that the velocity of each inclusion is a linear function of the applied forces. The velocity of the first inclusion, for example, is given by

\begin{equation}
{\bf V}_1 = {\bf M}_{11} {\bf F}_1 + {\bf  M}_{12} {\bf F}_2,
\label{eq:forces}
\end{equation}

\noindent
where ${\bf M}_{11}$ is the self-mobility matrix and ${\bf M}_{12}$ the mutual mobility matrix.  Since a pair of circular inclusions has axial symmetry, the only  non-vanishing components of the mobility matrices are the diagonal elements $M_{11}^{rr}$, $M_{11}^{\theta\theta}$, $M_{12}^{rr}$, and $M_{12}^{\theta\theta}$~\cite{kimbook}, where $rr$ refers to the radial motion of the inclusions (along the line connecting their centers), and $\theta \theta$ to tangential motion (perpendicular to this line). The mutual mobilities can be extracted from the experimental data by computing the cross-correlation function~\cite{Crocker2000}

\begin{eqnarray}
\lefteqn{\langle \Delta {\bf r}_1 (t) \cdot \Delta {\bf r}_2(t) \delta( r_{12}(0) - s) \rangle = }\nonumber\\
&&\qquad 2 k_B T ( M_{12}^{rr}(s) +  M_{12}^{\theta\theta}(s)  )  \, t \; ,
\label{eq:cross-corr-fn}
\end{eqnarray}

\noindent
where  $\Delta\mathbf{r}_k(t)=\mathbf{r}_k(t)-\mathbf{r}_k(0)$ is the displacement of the $k$th inclusion in time interval $t$, and $r_{12}(0)$ and $s$ are respectively the distances between the centers of the inclusions  at $t=0$ and at time $t$ (see Fig.~\ref{fig:island}).

\begin{figure}[tbh]
\begin{center}
\includegraphics[width=7cm]{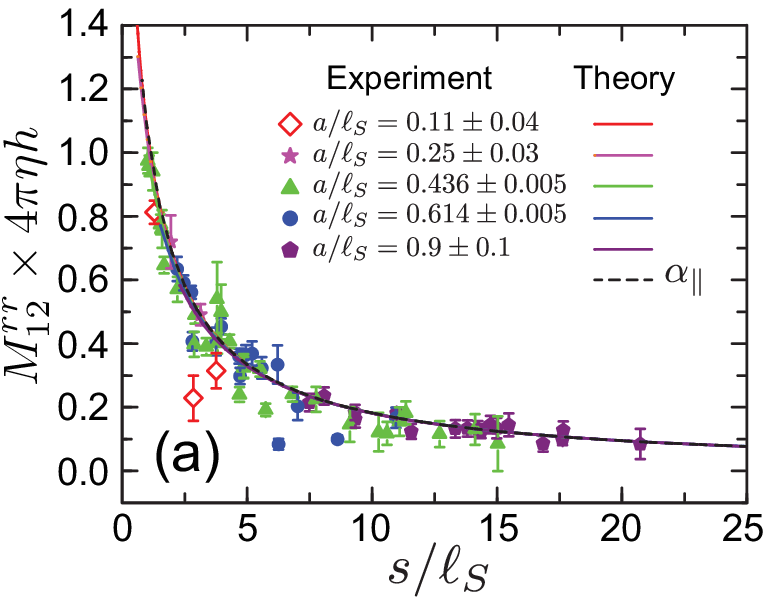}
\includegraphics[width=7cm]{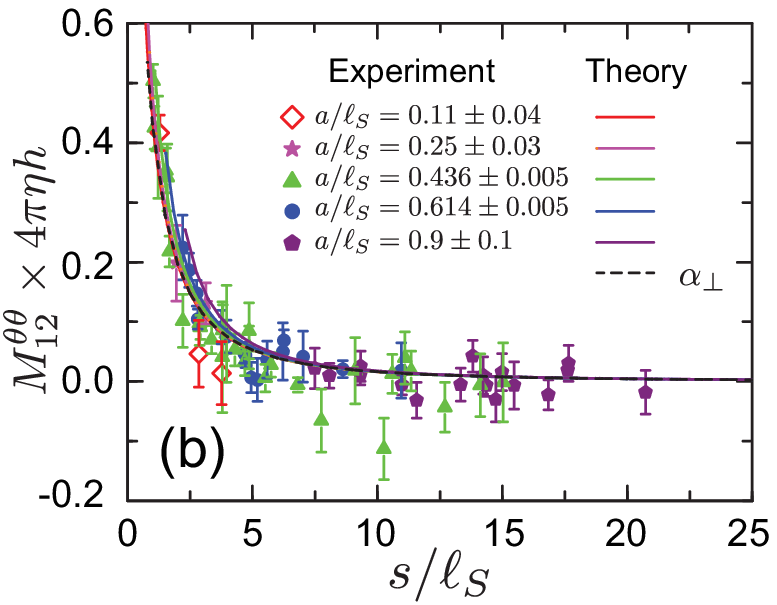}
\caption{(Color online) Measured and calculated mutual mobilities (a) $M^{rr}_{12}$ (radial) and (b) $M^{\theta\theta}_{12}$ (tangential) of pairs of oil droplets with radii  $a\approx b <\ell_S$ in smectic membranes as a function of dimensionless separation $s/\ell_S$. The statistical uncertainties here and in Fig.~\ref{fig:islands_results} are a consequence of combining measurements on several droplet pairs of the same average size. The LM response functions $\alpha_\|$ and $\alpha_\perp$ are shown as dashed curves.}
\label{fig:droplets_results}
\end{center}
\end{figure}

The observed dependence of the mutual mobilities $M_{12}^{rr}$ and $M_{12}^{\theta \theta}$ (scaled by $4\pi\eta h$) on the dimensionless center-to-center distance $s/\ell_S$  are plotted in Fig.~\ref{fig:droplets_results} for pairs of oil drops with approximately equal radius $a \approx b <\ell_S$.
Remarkably, in this regime, where dissipation occurs primarily in the smectic film~\cite{Stone2010}, the mutual mobilities are found to be independent of the inclusion size. Even when the drop separation $s$ is comparable to or smaller than the Saffman length, both $M_{12}^{rr}$ and $M_{12}^{\theta \theta}$ are independent of radius and closely follow the $\alpha_\|$ and $\alpha_\perp$ components of the response function tensor $\alpha^{\alpha\beta}$ derived by Levine and MacKintosh (black dashed lines in Fig.~\ref{fig:droplets_results})~\cite{Levine_MacKintosh2002}. In the LM theory, the response function $\alpha^{\alpha\beta}$ gives the flow induced by a
point force at ${\bf x}$:
$v^\alpha ({\bf x}') = \alpha^{\alpha\beta}({\bf x}' - {\bf x}) f^\beta \delta({\bf x})$.
Since the LM model describes the microrheology of visco-elastic membranes and we treat the smectic films as purely viscous 2D fluids, we use an
$\bm{\alpha}$ corresponding to  $-i\omega\bm{\alpha}$ in their theory. The response function $\alpha^{\alpha\beta}$ may be split into parallel (radial) and perpendicular (tangential) contributions,
$\alpha^{\alpha\beta}({\bf x}) = \alpha_\|(z) \hat{x}^\alpha \hat{x}^\beta + \alpha_\perp(z)
[\delta^{\alpha\beta} - \hat{x}^\alpha\hat{x}^\beta]$, where $z=|{\bf x}|/\ell_S$. Both $\alpha_\|(z)$ and $\alpha_\perp(z)$ diverge logarithmically as $z\rightarrow0$, and for large $z$ we have $\alpha_\|(z)\sim1/z$ and  $\alpha_\perp(z) \sim 1/z^2$.

\begin{figure}[tbh]
\centering
\includegraphics[width=7cm]{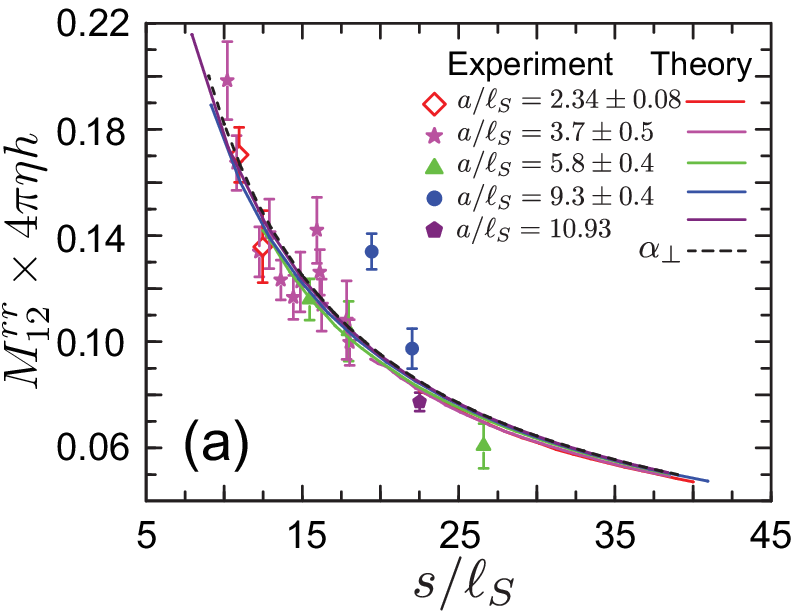}
\includegraphics[width=7cm]{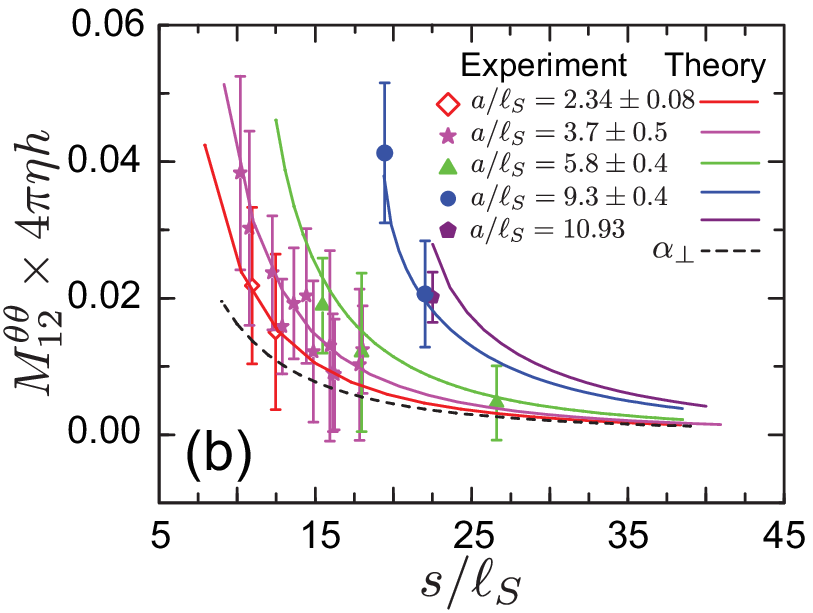}
\caption{(Color online) Measured and calculated mutual mobilities (a) $M^{rr}_{12}$ (radial) and (b) $M^{\theta\theta}_{12}$ (tangential) of pairs of islands with radii  $a\approx b > \ell_S$ in smectic membranes as a function of dimensionless separation $s/\ell_S$. The LM response functions $\alpha_\|$ and $\alpha_\perp$ are shown as dashed curves.}
\label{fig:islands_results}
\end{figure}

When we observe large smectic islands with $a \approx b > \ell_S$, the hydrodynamic regime in which dissipation occurs primarily in the air surrounding the smectic film, the radial mutual mobility $M^{rr}_{12}$ is described well by the LM response function $\alpha_\|$ for point-like particles even when the inclusions are very close together (Fig.~\ref{fig:islands_results}a).
The tangential mutual mobilities $M^{\theta\theta}_{12}$, on the other hand, deviate significantly from $\alpha_\perp$ and depend strongly on radius (Fig.~\ref{fig:islands_results}b).
When the inclusions in this regime have different sizes, however, the LM theory reproduces neither the radial nor the tangential mobilities observed experimentally.


\begin{figure}[tbh]
\centering
\includegraphics[width=7cm]{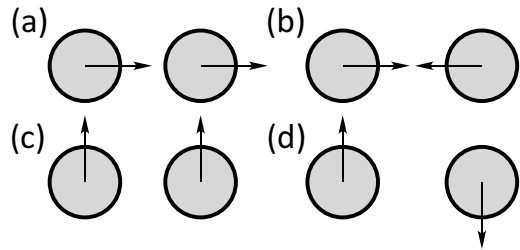}
\caption{Configurations of membrane inclusions moving in response to applied forces. The arrows represent both the applied forces (which all have the same magnitude) and the resultant velocities of the inclusion centers. The fluid is assumed to be at rest at infinity.
Determination of the viscous drag force using Eq.~\ref{eq:forces} in each case yields the following mobility combinations:
(a) $M^{rr}_{11} + M^{rr}_{12}$, (b) $M^{rr}_{11} - M^{rr}_{12}$, (c) $M^{\theta \theta}_{11} + M^{\theta \theta}_{12}$, (d) $M_{11}^{\theta\theta} - M_{12}^{\theta\theta}$.
}
\label{fig:islands_configurations}
\end{figure}

These observations confirm that at the crossover from 2D to 3D hydrodynamics (when $a,b \agt \ell_S$), the mutual mobilities depend in general both on the size of the inclusions and the distance between them.  This motivated us to extend the LM model beyond the far-field approximation in order to be able to characterize the interactions of circular inclusions of arbitrary radius.  We begin by calculating the flow field in the film generated by the motion of an inclusion of radius $a$, integrating the effects of an array of point forces along the inclusion boundary:

\begin{eqnarray}
\label{eq:membrane-velocity-dueto-island}
v^\alpha ( {\bf x}' ) &=& \sum_{j=1,2} \int_0^{2\pi}\!\!  d \phi\, f_j^\beta(\phi)  \alpha^{\alpha\beta}({\bf x}' - {\bf x}_j(\phi)) \; ,
\end{eqnarray}

\noindent
where $\alpha,\beta = x, y$; $j=1,2$ labels the inclusion; $f_j^\beta(\phi) $ are the (initially unknown) strengths of the point forces; and  $\alpha^{\alpha\beta}$ is the LM response function.  The force densities $f^\beta_j(\phi)$ are found by demanding no-slip boundary conditions $v^\alpha=V_j^\alpha$ at each inclusion and numerically solving Eq.~\ref{eq:membrane-velocity-dueto-island}.
In the particular case of two inclusions of equal radius, moving with velocities $V$ and subject to forces of equal magnitude $F$, one may determine the self- and mutual mobilities by invoking the linearity of the governing equations (Eq.~\ref{eq:forces}) and considering the four special inclusion configurations shown in Fig.~\ref{fig:islands_configurations}.

The mutual mobilities obtained in this way (solid curves in Figs.~\ref{fig:droplets_results} and \ref{fig:islands_results}) fit the experimental data well for all experimentally accessible inclusion sizes and ratios of $a/\ell_S$.  Our model predicts further that the self-mobilities $M_{kk}^{rr}$ and $M_{kk}^{\theta\theta}$ depend on the distance between the inclusions,  being reduced when another inclusion is nearby. However, this is a relatively weak effect that is difficult to measure in our experiments. As additional verification of the validity of our approach, we note that when applied to calculating the (self-)mobility of isolated inclusions with radii in the range $0.1 \ell_S < a < 10\ell_S$, our theory matches the HPW predictions~\cite{HPW1981} very well, confirming that accurate results can be obtained by assuming that the inclusions are rigid~\cite{Nguyen2010} and ignoring their interiors, modeling the total force on each inclusion as the superposition of the forces along its circumference.

In summary, we have probed the hydrodynamic interactions of a pair of inclusions in a thin fluid membrane. A theory generalizing the point particle approach of Levine and MacKintosh in order to consider inclusions of finite extent and arbitrary separation reproduces the experimental mobilities obtained by measuring the Brownian motion of oil droplets and islands in thin smectic~A films. The model confirms the surprising experimental observation that the mutual radial mobilities are independent of size for all Saffman lengths, while
the mutual tangential mobilities depend strongly on both size and separation only when the inclusions are larger than the Saffman length.

Using the theoretical tools developed here to describe pairs of inclusions in a model $2$D system, we plan to extend our studies to consider multiple inclusions, both mobile and immobile, in fluid films in order to better mimic the conditions found in cellular membranes. A possible application of our approach is to $2$D aggregation: while it has been shown that hydrodynamic interactions hinder aggregation in three dimensions~\cite{Deutch1973}, there has been little work on this effect in membranes.

We thank Jixia Dai for his generous assistance with the vacuum system and Joe Becker for his help in setting up the experiment. This work was supported by NASA Grant~NNX-13AQ81G and NSF MRSEC Grant~DMR-0820579 (University of Colorado), and by NSF Grant~CBET-0854108 (Brown University).


\bibliographystyle{apsrev}
\bibliography{smecticislands} 

\end{document}